\documentstyle[prl,aps,floats]{revtex}

\begin{document}
\preprint{astro-ph/9906327}
\draft

%
%
\input epsf

\renewcommand{\topfraction}{0.99}
\renewcommand{\bottomfraction}{0.99}

\twocolumn[\hsize\textwidth\columnwidth\hsize\csname
@twocolumnfalse\endcsname

\title{Inflaton potential reconstruction without slow-roll}
\author{Ian J.~Grivell and Andrew R.~Liddle}
\address{Astrophysics Group, The Blackett Laboratory, Imperial College, London 
SW7 2BZ, United Kingdom} 
\date{\today} 
\maketitle
\begin{abstract}
We describe a method of obtaining the inflationary potential from
observations which does not use the slow-roll approximation. Rather, the
microwave anisotropy spectrum is obtained directly from a parametrized
potential numerically, with no approximation beyond linear perturbation
theory. This permits unbiased estimation of the parameters describing
the potential, as well as providing the full error covariance matrix. We
illustrate the typical uncertainties obtained using the Fisher
information matrix technique, studying the $\lambda \phi^4$ potential in detail 
as a concrete example.
\end{abstract}

\pacs{PACS numbers: 98.80.Cq, 98.70.Vc \hspace*{6.3cm} astro-ph/9906327}

\vskip2pc]

\section{Introduction}

The determination of the initial power spectrum of perturbations in the
Universe is an essential step in using the cosmic microwave background
to constrain cosmological parameters. Indeed, without an understanding
of these initial perturbations, the cosmic microwave background in
isolation says nothing about the values of parameters such as the Hubble
parameter $h$ and the density parameter $\Omega_0$. The reason is that
the effect of the cosmology is on the {\em dynamics} of the
perturbations, and a single time-slice, such as the perturbations at
last scattering, says nothing about the dynamics. Usually, this problem
is circumvented by assuming a parametrization of the initial conditions;
the cosmological parameters then enter via the dynamics converting these
initial conditions into the conditions at last scattering. For example,
a common assumption is that there is a power-law spectrum of gaussian
adiabatic scalar perturbations. This is a popular choice because it fits
current observations, and because it is a good approximation to the
perturbations produced by the simplest inflation models \cite{genrefs,LL}, but 
is clearly
quite specific since it requires four descriptive
qualifiers.\footnote{In fact, as shortly discussed, there's actually a
fifth qualifier as it is normally assumed that these perturbations are
entirely in the growing mode.}

Given a set of microwave anisotropy measurements, one must determine
both the cosmological parameters and the parameters describing the
initial perturbations simultaneously; it is not possible to do one and
not the other. Given the description of the initial perturbations, one
can then try to determine the model which gave rise to them. If the
best-fit model proves to have passive perturbations, meaning that the
perturbations are observed to be entirely in their growing mode and
hence are inferred to have existed since early in the Universe's
evolution (in particular, to have already existed when their scale was
considerably larger than the Hubble radius), then it is reasonable to
believe that they arose via the inflationary mechanism \cite{L95,HW},
rather than being induced by topological defects or other causal mechanism. It 
would be further
encouraging if the perturbations proved to be adiabatic and gaussian,
because although inflation models exist which violate those conditions,
they are properties of the simplest inflation models.

Attempting to derive the underlying inflation model from observations
has become known as inflaton potential reconstruction \cite{recon,LLKCBA}. 
Studies have
focussed on models where inflation is driven by a single scalar field
$\phi$, moving in a potential $V(\phi)$. Such models indeed give
perturbations which are passive, adiabatic and gaussian, though they
come in two types, scalar (density perturbations) and tensor
(gravitational waves) which need not be perfect power-laws. Even then they are 
not the most general class of
models leading to that set of properties, because models where there is
more than one scalar field can also give rise to that outcome. However,
the single-field case appears to be the largest class of models which
can be dealt with as a single set, where one aims to identify the member
of the set responsible for the observations. If it turns out that there
is no such member, then the net must be cast wider to include more
complicated models. Amongst more general inflation models, there is no problem 
in
generating the required predictions of the spectra to test them one by
one against the data, but there is no known way of looking at the
observations and constructing an inflationary model which will generate
the desired predictions. We stress again that unless a valid model for
the initial perturbations can be found, one cannot obtain the
cosmological parameters as there would be no way to compute a microwave
anisotropy spectrum to compare with observations.

Even under the single-field paradigm, the situation has not been wholely
satisfactory, the reason being that the analytic results available for
the spectra are only approximate, having been calculated using the
slow-roll approximation, within which results are known only to
second-order. Therefore, even if unbiased estimates of the parameters
describing the perturbation spectra are obtained from observations (such
as the amplitude and the spectral index $n$), these are not translated
into unbiased estimators of the inflation potential. This paper describes how 
this shortcoming can be overcome.

\section{Direct estimation of inflationary parameters}

Throughout, we work within the single field paradigm. The traditional
technique for obtaining observational predictions from an inflationary
model is the following. The potential is specified as an analytical
function. The perturbations are then computed using the slow-roll
approximation, to give the perturbation spectra in parametrized form.
For example, the density perturbation spectrum $\delta_{{\rm H}}(k)$
(following the notation of Liddle and Lyth \cite{LL}) can be expanded as
a Taylor series in $\ln k$ as
\begin{eqnarray}
\label{running}
\ln \delta_{{\rm H}}^2(k) & = & \ln \delta_{{\rm H}}^2(k_*) +
    (n_*-1) \, \ln \frac{k}{k_*} + \nonumber \\
 & & \quad \frac{1}{2} \left. \frac{dn}{d\ln k} 
    \right|_{k_*} \, \ln^2 \frac{k}{k_*} + \cdots \,,
\end{eqnarray}
where $k$ is the comoving wavenumber and $k_*$ is an arbitrary scale
where the coefficients are evaluated. The slow-roll approximation gives
the coefficients as functions of the potential, but only approximately, and 
there is a further
approximation when the series is truncated at some level. Further,
because the expression relating the scalar field value $\phi$ to the
scale $k$ crossing the horizon is also approximate, there is a problem
of a `drift of scales'; as we move from the expansion scale $k_*$, we begin to
misidentify the $\phi$ value corresponding to a $k$ value by more and
more. 

Often all these errors are unimportant, especially for observations of
the current quality. That for certain models they give an 
error which will be significant for future observations has been noted by 
several authors \cite{Wangetal,cgkl,ms}. One can
attempt to improve things by going to the highest possible order in the
slow-roll expansion, which is unfortunately only second-order, or by
taking more and more terms in  Eq.~(\ref{running}) \cite{KosT,cgl},
which does not require going to higher order in slow-roll. 

One does the best one can with the scalar perturbations, and also
carries out a similar process for the tensor spectrum, which is less
demanding theoretically as tensors are harder to detect observationally. These
parametrized spectra are then fed into a numerical code (e.g.~{\sc
cmbfast} \cite{SZ}, possibly enhanced to allow non-power-law spectra) to
compute the microwave anisotropy spectrum, the $C_\ell$, as a function of those
and the cosmological parameters.

In going in the reverse direction, starting with observations, the standard
procedure is to use the observations to estimate the coefficients in
Eq.~(\ref{running}), along with the cosmological parameters. An example using 
current data is the analysis by Tegmark \cite{Teg8}.
As long as
the parametrization of the perturbations was adequate, that's the job
done as far as the cosmological parameters are concerned. However, to
obtain the inflationary potential, the {\em approximate} slow-roll results
which give the spectra in terms of the inflationary potential are
inverted. This procedure is reviewed in Ref.~\cite{LLKCBA}, but does not
yield unbiased estimates of the inflation potential. 

The numerical technology now exists to circumvent this
problem. The key is to immediately abandon any attempt to make the
calculations analytically. Instead, the perturbation spectra are obtained
by numerical solution of the relevant mode equations, which give the
perturbation amplitude at a particular wavenumber. The best formalism
is that of Mukhanov \cite{Muk}, and the only assumption is that linear
perturbation theory is valid, which is more or less guaranteed by the
fact that the observed (dimensionless) perturbations are order
$10^{-5}$. The necessary ingredients to proceed are
\begin{enumerate}
\item A program which can numerically solve the mode equations wavenumber by 
wavenumber. We
described such a code in Ref.~\cite{gl}. This must be able to compute
both scalar and tensor perturbations.
\item A version of {\sc cmbfast} which is capable of taking
arbitrary power spectra as input to produce the $C_\ell$ curve. The
output $C_\ell$ is the sum of the scalar and tensor parts. Polarization 
anisotropies should be computed as well as temperature ones.
\end{enumerate}
We have assembled these codes into an IDL pipeline. The input step is
to supply a parametrization of the potential, rather than an
analytic form. In this paper we use the simplest version, a Taylor 
expansion about some
scalar field value $\phi_*$, with the slight subtlety of pulling out the overall 
normalization as a prefactor for later convenience. This leads to the $C_\ell$ 
as a function of
the cosmological parameters and the potential parameters, 
i.e.~$C_\ell(V_*,V_*'/V_*,V_*''/V_*,...,h,\Omega_0,\Omega_{{\rm
B}},\Omega_\Lambda,...)$ where primes are derivatives with respect
to $\phi$, evaluated at $\phi_*$.

The inversion is now direct; the observed anisotropy spectrum is used to
directly estimate the potential parameters, which can be done in an
unbiased way to generate the best possible reconstruction. If at this
stage one were to find that the overall best-fit model was a poor fit to
the data, the first thing would be to try an improved parametrization of
the potential and/or inclusion of extra cosmological parameters, and if that 
still fails it would be time to suspect that
the single field paradigm is not correct. 

However, optimistically assuming that the best fit is adequate, we have
our best-fit inflationary potential. But that's not all; a further
advantage of this direct method is that it immediately gives us the
covariance of the uncertainties on the potential parameters. For
example, it is known that the errors on $V_*$ and $V_*'$ will be highly
correlated. Using the old approach, these correlations would have to be
carried through the complicated reconstruction equations, an unpleasant
enough task that in Ref.~\cite{cgkl} we instead used a Monte Carlo
method to illustrate the uncertainties of a reconstruction from simulated
data. 

In this approach, the consistency equation relating scalar and tensor 
perturbations (see Ref.~\cite{LLKCBA} for a discussion) is automatically 
incorporated, being tested by whether there is a potential offering a 
satisfactory absolute fit to the data. It could of course also be tested in the 
traditional way by power spectrum fitting, but anyway it is unlikely that 
observations will be good enough to say anything significant.

\begin{figure}
\centering 
\leavevmode\epsfysize=4.4cm \epsfbox{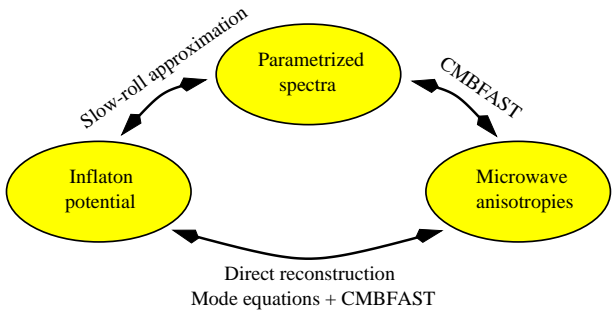}\\ 
\caption[strategy]{\label{f:strategy} The traditional route from model to 
observables and back is the two-stage process at the top. The procedure outlined 
in this paper enables a direct route without approximations beyond linear 
perturbation theory.}
\end{figure}

The two strategies are contrasted in Fig.~1. We do not view our new approach as 
replacing the traditional one, but rather as a next step that one would take if 
the traditional fitting proves successful, in order to obtain optimal results.

\section{Uncertainty and covariance of inflaton potential parameters}

\subsection{Parametrizing the potential}

Although in principle a solution of the mode equation runs from an early
initial time until the scale is well outside the horizon, when
considering perturbations on a given scale $k$, the details of how the
Universe expands are only important for a fairly brief interval around
the time $k=aH$ when the scale crosses outside the horizon. The reason
is that while the scale is well inside the horizon the relevant
timescales are much less than the expansion timescale and expansion can
be neglected, while when scales are above the horizon the perturbations
are frozen in at fixed values (in the appropriate variables) whatever
the evolution is. As the observations cover a limited range of
wavenumbers $k$, we need only know $V(\phi)$ for a limited range of
$\phi$ values about the time when the relevant scales cross outside the
horizon during inflation. Our input assumption (ultimately to be tested
against the observations) is that there is a potential $V(\phi)$, and we
will simply need a parametrization of it which is accurate enough over
the desired range.

There is one subtlety to this. The scalar wave equation is second order,
so, in addition to the value of $\phi$, it looks as if $\dot{\phi}$ is an
arbitrary initial condition which needs to be considered as an extra
parameter. However it has long been known that this is not the case,
because scalar field cosmologies have an attractor behaviour whereby all initial
conditions quickly converge \cite{SB,LPB} (indeed, during inflation convergence 
is at
least exponentially fast with $\phi$, but even non-inflationary expansion 
exhibits this behaviour). However, this does
mean that we have to be sure that the simulation has run for long enough
that the attractor is attained before the perturbations on observable
scales are generated, exactly as is believed to have happened in the real
Universe. 

As in known inflation models all observable scales cross outside the
horizon over a very narrow range of $\phi$, the simplest approach is a
simple Taylor series expansion
\begin{equation}
\frac{V(\phi)}{V_*} = 1 + \frac{V'_*}{V_*} \, (\phi - \phi_*) + 
\frac{1}{2} \frac{V''_*}{V_*} \, (\phi-\phi_*)^2 + \cdots  \,,
\end{equation}
where $\phi_*$ is arbitrary and can be set to zero if desired. We pulled out the 
normalization before expanding, as then the normalization of the $C_\ell$ 
depends only on $V_*$ and not the other terms.
In principle one could consider a more sophisticated expansion to try
and improve the convergence properties such as a Pad\'e approximant, but
that can be assessed once actual data is available.

\subsection{Parameter uncertainty}

Having obtained the $C_\ell$ as a function of the potential and
cosmological parameters, we can assess the likely
accuracy with which those parameters can be found by a given experiment.
This is carried out using the well-established Fisher matrix technique
\cite{parest,ZSS}, which amounts to taking the derivative of the $C_\ell$
with respect to each of the parameters. The parameter uncertainties
depend on the choice of `correct' model and on the number of
parameters allowed to vary. For illustration, we vary cosmological
parameters about an underlying model with Hubble parameter $h = 0.65$, density
parameter $\Omega_0 = 0.3$, cosmological constant $\Omega_\Lambda =
0.7$, baryon density $\Omega_{{\rm B}} = 0.05$ and reionization
optical depth $\tau = 0.1$. 

The potential we choose is the $\lambda \phi^4$ potential, and we take the epoch 
where the present Hubble radius equaled the Hubble radius during inflation as 
being 60 $e$-foldings before the end of inflation. Numerical 
solution of the equations of motion gives this as $\phi_* = 4.37 \, m_{{\rm 
Pl}}$. In fact the slow-roll approximation will work well for this potential, 
and for instance can be used to show that gravitational waves should contribute 
about twenty percent of the signal at large angular scales.

Following Zaldarriaga et al.~\cite{ZSS}, we consider a version of the {\sc 
Planck} satellite which measures both
temperature and polarization anisotro\-pies, as described in Ref.~\cite{cgl}.
We do not attempt to include the effects of foregrounds, as extensively studied
recently by Tegmark et al.~\cite{TEHO}, but choose to consider only one
polarized Planck channel (in effect assuming that the polarized foregrounds can
be removed using all the other channels) which, while rather approximate, yields
similar results.  The actual data, when available, will of course merit more
sophisticated treatment. Our numbers are therefore indicative only, and more 
importantly they would vary significantly if the assumed underlying model were 
changed --- the quality of information available from reconstruction depends 
strongly on which model (if any, of course) proves to be correct. 

The results are shown in Table~I. The higher derivatives are not detected, but 
it is interesting to note that even with this very flat potential, the variation 
of the potential during inflation, $V'$, is detected at 7-sigma. However $V''$ 
is not detected; this may seem a little surprising given that both the 
gravitational wave amplitude and the scalar spectral index (which depend on 
different combinations of $V'/V$ and $V''/V$) are in fact detectable for this 
potential \cite{parest,ZSS,cgl,TEHO}, 
but it turns out that the combination of these giving $V''/V$ is not 
distinguishable from zero. In our approach one never needs to make the 
separation of scalars and tensors explicitly.

\begin{table}
\begin{tabular}{ccc}
&\multicolumn{1}{c}{underlying} &\multicolumn{1}{c}{relative}\\
parameter &\multicolumn{1}{c}{model value} &\multicolumn{1}{c}{uncertainty}\\
\hline
$\tau$ &0.1 &6.1\% \\
$\Omega_{\rm B}h^2$ &0.021 &1.2\% \\
$\Omega_{\rm CDM}h^2$ &0.11 &2\% \\
$\Omega_{\Lambda}h^2$ &0.30 &5\% \\
&& \\
$10^{12} \, V_*/m_{{\rm Pl}}^4$ &$2.3 $ &22\% \\
$m_{{\rm Pl}}\,V'_*/V_*$ &$0.92$ &14\% \\
$m_{{\rm Pl}}^2\,V''_*/V_*$ &$0.63$ &$2\times$ \\
$m_{{\rm Pl}}^3\,V'''_*/V_*$ &$0.29$ &$60\times$ \\
$m_{{\rm Pl}}^4\,V''''_*/V_*$ &$0.066$ &$400\times$\\
\end{tabular}
\caption{Uncertainties for each parameter, marginalizing over the
remaining parameters. We stress that these are specifically for the $\lambda 
\phi^4$ model. These values correspond
to the diagonal entries of the covariance matrix. There are
substantial correlations between parameters, especially those
describing the potential, so the off-diagonal entries of the
covariance matrix are significant. Of the potential parameters, only the 
magnitude and gradient of the potential are detected with any significance.}
\end{table}

\subsection{Parameter uncertainty covariance}

The Fisher matrix technique also generates the covariances of the error
estimates, and these are crucial in interpreting the observational constraints. 
In particular, the covariance matrix is essential in illustrating the 
reconstructions graphically; if correlations are ignored then the reconstruction 
deteriorates much more quickly with $\phi$ than the true picture (the 
correlations allow for the fact that scale of the expansion $\phi_*$ need not be 
the scale at which the observations are the most powerful). That our method 
gives the full correlation matrix of the reconstructed potential directly is its 
first key advantage over earlier techniques.

\begin{figure}[t]
\centering 
\leavevmode\epsfysize=6.25cm \epsfbox{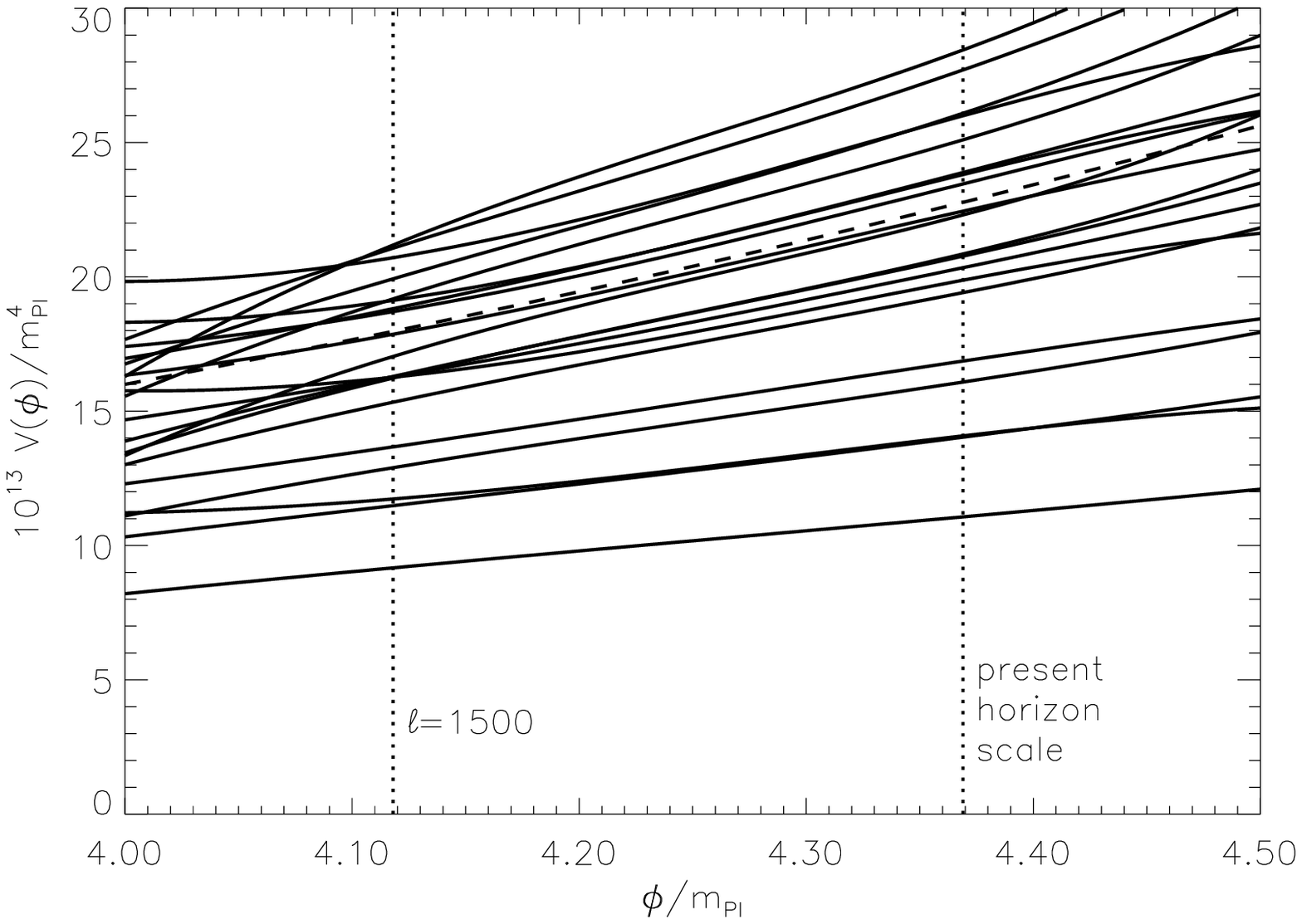}\\ 
\leavevmode\epsfysize=6.25cm \epsfbox{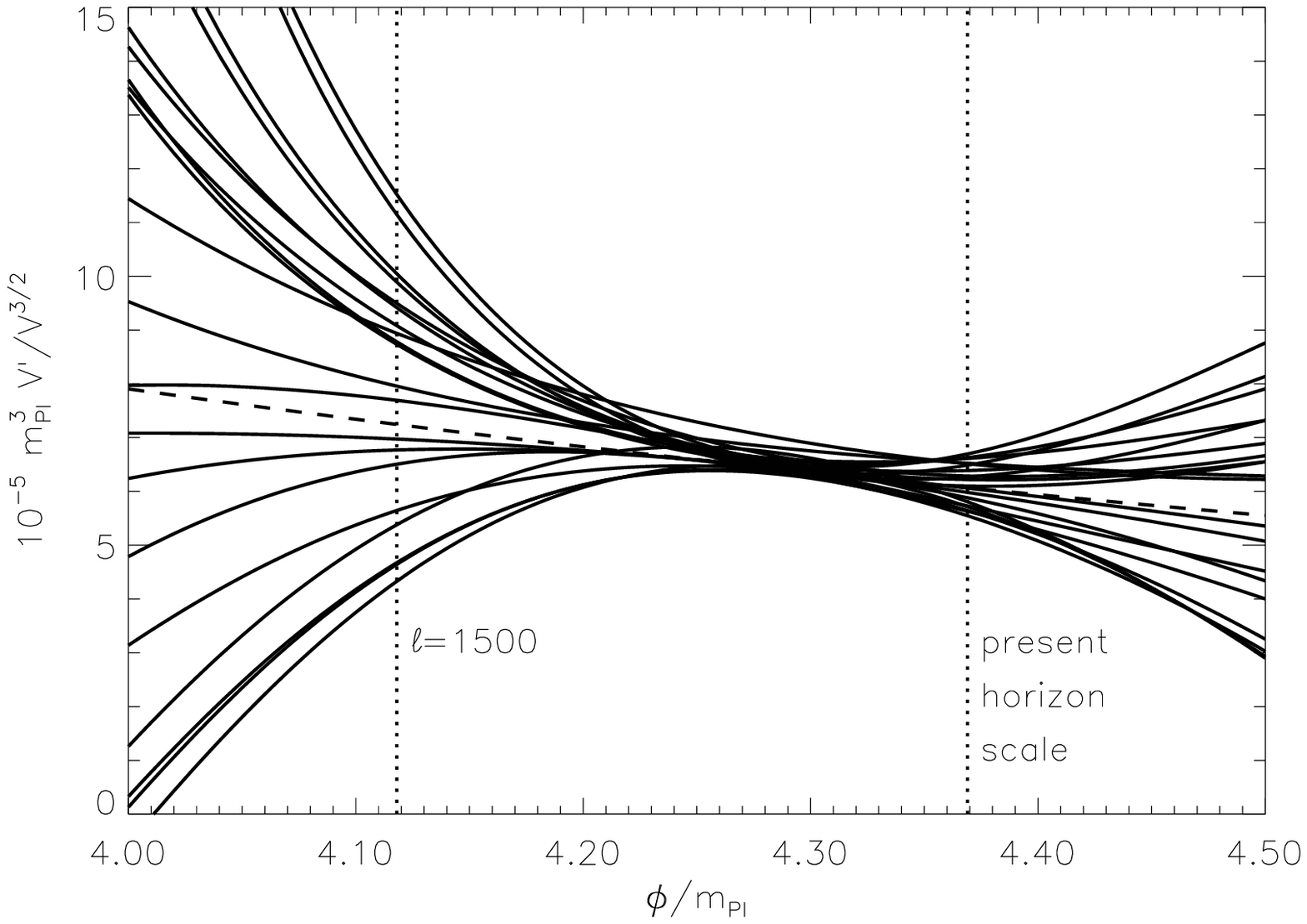}\\ 
\caption[recons]{\label{f:recons} Twenty Monte Carlo reconstructions of the 
potential, 
compared against the true potential which is shown as a dashed line. The upper 
panel shows the potential itself, and the lower one the combination $V'/V^{3/2}$ 
which is a combination coming primarily from the density perturbations alone. 
The dotted vertical lines indicate the region of the potential directly probed 
by the microwave background, ranging from the current horizon scale to the 
horizon scale when the $\ell = 1500$ mode was generated (evaluated in the 
underlying model). The upper panel shows that the 
gradient is quite well recovered but the overall amplitude much less so, while 
the lower highlights the obvious fact that the reconstruction is accurate only 
where there is data available to constrain it.}
\end{figure}

To illustrate the quality of the reconstruction, we carry out Monte Carlo 
reconstructions with errors drawn according to the covariance matrix, and plot 
them against the true potential in Fig.~\ref{f:recons}. These reconstructions 
include up to the fourth derivative; though as seen from Table~I the higher 
derivatives are not detected, they can still be assigned values according to 
their upper limit.
We note that this potential is much less favourable for reconstruction than ones 
explored previously, as it is much closer to the scale-invariant limit.

The reconstructions indicate the second key advantage of the method proposed 
here over previous ones (e.g.~Refs.~\cite{LLKCBA,cgkl}) --- the reconstructed 
potentials are unbiased estimates of the true potential, being as likely to be 
too high as too low. In the upper panel, we see that the uncertainty in the 
overall normalization is quite large (ultimately due to a degeneracy in the 
effect of 
scalars and tensors on large angular scales). The lower panel shows the 
combination $V'/V^{3/2}$ which is primarily sensitive to density perturbations 
alone (indeed perfectly so in the slow-roll approximation), and which is much 
better determined (in particular, better than $V$ or $V'$ separately). This 
figure allows us to see directly the range of $\phi$ 
which is constrained by the data; to highlight this we have indicated the values 
which $\phi$ takes while the microwave anisotropies are being generated. 
At larger $\phi$, corresponding to larger scales, the perturbations are 
unobservable, and even some way within the current horizon scale cosmic variance 
contributes significantly to the spread. Near the centre of the data the 
determination is at its best, and on short scales the information again becomes 
poor, partly because of the dependence on all the cosmological parameters and 
partly because Silk damping erases the perturbations as one goes 
beyond $\ell \sim 1000$.

This figure highlights once again that the information available from 
reconstruction constrains only a tiny portion of the scalar field potential. 
Nevertheless, the information available there is of good accuracy, and can 
be highly constraining in instances where theoretical motivation suggests a 
potential containing few unspecified parameters.

\begin{figure}[t]
\centering 
\leavevmode\epsfysize=6.25cm \epsfbox{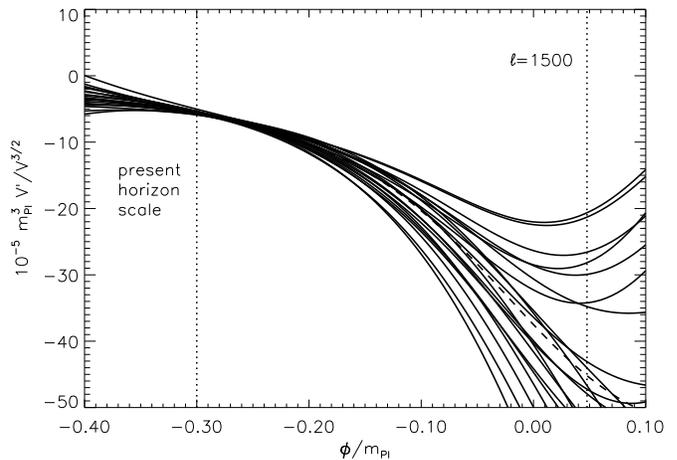}\\ 
\caption[wang]{\label{f:wang} Twenty Monte Carlo reconstructions of
the combination $V'/V^{3/2}$, as in the lower panel of
Fig.~\ref{f:recons}, but for the model investigated by Wang et
al.~\cite{Wangetal}.}
\end{figure}

The quartic potential is an interesting test case because it is not far from the 
slow-roll limit, and this is the first time such a potential has been used to 
test reconstruction methods. However the true strength of the method would be 
unveiled if the true model does not satisfy slow-roll well, despite the 
potential being smooth. An example is the potential introduced by Wang et 
al.~\cite{Wangetal}, which was used to test the traditional reconstruction 
technique in Ref.~\cite{cgkl}. It was shown in that latter paper that 
traditional reconstruction could still work well, but led to a bias in the 
estimate of the potential (albeit within observational
errors). Figure~\ref{f:wang} illustrates test reconstructions of this
potential, using the techniques of the present paper. The
uncertainties on the cosmological parameters, and on $V$ and $V'$, are
very similar to those of the quartic case. However, in addition $V''$
is detected at around 3-sigma and the next two derivatives have uncertainties
comparable to their values.

\section{Discussion}

It may well be that one of the simplest models of inflation is correct, and the
perturbation spectra are perfectly satisfactorily approximated by a power-law
[or at least some low-order truncation of Eq.~(\ref{running})].  If so, then
cosmological parameter estimation can proceed as described in previous works.
However, given the intellectual and financial investment in pursuing 
cosmological parameters, it is vital to
be aware of possible difficulties, and to analyze ways of dealing with them.  We
have considered one such possibility --- that the single field paradigm is
correct but slow-roll is not very good --- and explained that this is readily
dealt with using existing numerical technology. In using this technique, as with 
others, it is imperative to make an overall goodness-of-fit test to ensure that 
the class of models being considered is capable of adequately explaining the 
data. Even if the power-law
approximation proves valid (and certainly this is the method which should be 
tried first), one will want to use these techniques to ensure that
estimates of the inflationary potential are unbiased ones, and to obtain the
fullest possible information about the inflaton potential from observations.

\section*{Acknowledgments}

I.J.G.~was supported by PPARC. We thank Ed Cope\-land and Rocky Kolb for many 
useful discussions and comments.
 

\begin{references}
\bibitem{genrefs} A. H. Guth, Phys. Rev. D {\bf 23}, 347 (1981);
	A. D. Linde, {\em Particle Physics and Cosmology}
	(Harwood, Chur, Switzerland, 1990).
\bibitem{LL} A. R. Liddle and D. H. Lyth, Phys. Rep. {\bf 231}, 1 (1993).
\bibitem{L95} A. R. Liddle, Phys. Rev. D {\bf 51}, 5347 (1995).
\bibitem{HW} W. Hu and M. White, Phys. Rev. Lett. {\bf 77}, 1687 (1996).
\bibitem{recon} E. J. Copeland, E. W. Kolb, A. R. Liddle and J. E. Lidsey,
	Phys. Rev. Lett. {\bf 71}, 219 (1993), Phys. Rev. D {\bf 48}, 2529
	(1993); M. S. Turner, Phys. Rev. D {\bf 48}, 5539 (1993).
\bibitem{LLKCBA} J. E. Lidsey, A. R. Liddle, E. W. Kolb, E. J. Copeland, T. 
	Barreiro and M. Abney, Rev. Mod. Phys. {\bf 69}, 373 (1997).
\bibitem{Wangetal} L. Wang, V. F. Mukhanov and P. J. Steinhardt,
	Phys. Lett. B {\bf 414}, 18 (1997).
\bibitem{cgkl} E. J. Copeland, I. J. Grivell, E. W. Kolb and A. R. Liddle,
	Phys. Rev. D {\bf 58}, 043002 (1998).
\bibitem{ms} J. Martin and D. Schwarz, {\tt astro-ph/9911225}.
\bibitem{KosT} A. Kosowsky and M. S. Turner, Phys. Rev. D {\bf 52}, 1739
	(1995).
\bibitem{cgl} E. J. Copeland, I. J. Grivell and A. R. Liddle, Mon. Not. 
	R. Astron. Soc. {\bf 298}, 1233 (1998).
\bibitem{SZ} U. Seljak and M. Zaldarriaga, Astrophys. J. {\bf 469}, 1
	 (1996).
\bibitem{Teg8} M. Tegmark, Astrophys. J {\bf 514}, L69 (1999).
\bibitem{Muk} V. F. Mukhanov, JETP {\bf 67}, 1297 (1988), Phys.
	Lett. B {\bf 218}, 17 (1989).
\bibitem{gl} I. J. Grivell and A. R. Liddle, Phys. Rev. D {\bf 54}, 7191
	(1996).
\bibitem{SB} D. Salopek and J. R. Bond, Phys. Rev. D {\bf 42}, 3936
	(1990).
\bibitem{LPB} A. R. Liddle, P. Parsons and J. D. Barrow, Phys. Rev. D
	{\bf 50}, 7222 (1994).
\bibitem{parest} L. Knox, Phys. Rev. D {\bf 52}, 4307 (1995); G. Jungman, M.
	Kamionkowski, A. Kosowsky and D. N. Spergel, Phys. Rev. D {\bf 54},
	1332 (1996); J. R. Bond, G. Efstathiou and M. Tegmark, Mon.
	Not. R. Astron. Soc. {\bf 291}, L33 (1997).
\bibitem{ZSS} M. Zaldarriaga, D. Spergel and U. Seljak, Astrophys. J. 
	{\bf 488}, 1 (1997).
\bibitem{TEHO} M. Tegmark, D. J. Eisenstein, W. Hu and A. de Oliveira-Costa,
	{\tt astro-ph/9905257}.
\end{references}
\end{document}